\documentclass[acmsmall]{acmart}
\AtBeginDocument{%
  }

\setcopyright{cc}
\setcctype{by}
\acmJournal{PACMHCI}
\acmYear{2025} \acmVolume{9} \acmNumber{2} \acmArticle{CSCW155} \acmMonth{4}
\acmPrice{}\acmDOI{10.1145/3711053}

\received{January 2024}
\received[revised]{July 2024}
\received[accepted]{October 2024}

\newcommand{\colored}{\textcolor{black}}
\begin{document}

\title{Reexamining Technological Support for Genealogy Research, Collaboration, and Education}

\author{Fei Shan}
\email{fshan@vt.edu}
\orcid{0000-0001-5070-4565}
\affiliation{%
  \department{Department of Computer Science \& Center of HCI}
  \institution{Virginia Tech}
  \city{Alexandria}
  \state{Virginia}
  \country{USA}
}

\author{Kurt Luther}
\email{kluther@vt.edu}
\orcid{0000-0003-1809-6269}
\affiliation{%
  \department{Department of Computer Science \& Center of HCI}
  \institution{Virginia Tech}
  \city{Alexandria}
  \state{Virginia}
  \country{USA}
}

\renewcommand{\shortauthors}{Fei Shan and Kurt Luther}

\begin{abstract}

Genealogy, the study of family history and lineage, has seen tremendous growth over the past decade, fueled by technological advances such as home DNA testing and mass digitization of historical records. However, HCI research on genealogy practices is nascent, with the most recent major studies predating this transformation. In this paper, we present a qualitative study of the current state of technological support for genealogy research, collaboration, and education. Through semi-structured interviews with 20 genealogists with diverse expertise, we report on current practices, challenges, and success stories around how genealogists conduct research, collaborate, and learn skills. We contrast the experiences of amateurs and experts, describe the emerging importance of standardization and professionalization of the field, and stress the critical role of computer systems in genealogy education. We bridge studies of sensemaking and information literacy through this empirical study on genealogy research practices, and conclude by discussing how genealogy presents a unique perspective through which to study collective sensemaking and education in online communities.

\end{abstract}




\maketitle

\section{Introduction}\label{sec:introduction}
Genealogy, the activity of studying family history and lineage, has a long history itself. Genealogists search for information regarding ancestors among surviving documents, discover history of a family through traces of individual family members, and organize articles and pedigree diagrams indicating parental relations of the family. Emerging genealogy websites such as Ancestry.com and FamilySearch.org has transformed traditional genealogy research since the early 2000s. These sites granted users the capability to access millions of digitized historical documents through the Internet, which attracted an increasing number of users who are interested in exploring histories of their families. Featuring the idea of discovering common ancestors and making connections to living relatives around the world, genealogy websites also allowed users to build family trees and connect them to others' trees, presenting a unique form of massive-scale online collaboration~\cite{de2022digital}.

These dramatic changes reverberated throughout the genealogy community, eliciting mixed reactions. In 1999, Elizabeth Shown Mills, president of the American Society of Genealogists and editor of \textit{National Genealogical Society Quarterly}, raised concerns that ``family history’s progress as an intellectually valuable pursuit'' could be undermined by ``web-grown genealogists'' who were ``largely unschooled in research principles'' but empowered by technologies to broadcast their genealogy work~\cite{mills1999working}. To ensure the quality of the genealogical research product, senior members of this community sought to formalize the desired research processes into guidelines and standards~\cite{ngs2018guidelines, bcg2023ethics}. 
However, these standards failed to gain traction outside of professional societies, and concerns of widespread genealogy research errors proved prescient. 

Indeed, as the spread of misinformation became a global concern, the genealogist community has faced its own version of this problem.  
A 2014 study~\cite{willever2014family} suggested that there was a considerable number of genealogy hobbyists, referred by serious genealogists as ``clickologists'' and ``half-researchers'', who built and connected to family trees on genealogy websites without knowing the proper genealogy research practices. Such behavior brought disturbances to the genealogist community: People found their close relative became a member of many strangers' families, and ``research hints,'' which are algorithmic recommendations of relevant information about an ancestor, would pick up inaccurate information from family trees and mislead more users. In recent years, an explosion of interest in genealogy has only intensified the problem.
As of 2019, FamilySearch claimed to be the home of the world's largest family tree, built collectively by its users and containing more than 1.2 billion ancestors \cite{familysearch2019world}.

Surprisingly, there has been minimal follow-up research from an human-computer interaction (HCI) or CSCW perspective on this challenge in the past decade. Recognizing the extensive growth of genealogical business fueled by the advancement of new technologies such as customized search engine~\cite{jiang2019ranking}, genealogical data visualization and analysis~\cite{liu2017genealogyvis, borges2019contextual, burch2021famsearch, wang2023framework}, DNA testing \cite{nash2004genetic, regalado2019more, calafell2017chromosome, stallard2020things, glynn2022bridging}, and artificial intelligence (AI) based handwriting recognition~\cite{nion2013handwritten, kennard2018computer, blomqvist2022joint}, face recognition~\cite{mohanty2020photo, robinson2021families} and family tree network exploration~\cite{malmi2017ancestryai, malmi2018computationally}, we believe there is an urgency to reexamine the current state of technology-supported genealogy research. In this paper, we attempt to fill the research gap by conducting a qualitative study of current genealogy practices. We extend prior work with new analysis comparing amateur and expert genealogists, investigating the impact of new technologies, and understanding how newcomers of this community learn to become competent researchers on the Internet. Specifically, our study is guided by the following research questions:

\begin{enumerate}
\item How has genealogists' research evolved with recent technology advances?
\item What are amateur genealogists' research practices compared with established standards?
\item How do amateur genealogists learn online, and how can technology better support this process?
\end{enumerate}

We conducted a semi-structured interview study with 20 genealogists, recruiting 10 experts with extensive professional research experience and 10 amateurs who are relatively new to this field. To the best of our knowledge, this study is the first that considers the role of standardization, professionalism, and credentials in the genealogist community, and to conduct a side-by-side comparison of research practices between genealogists with different levels of expertise. This work presents a more comprehensive description of genealogists' research processes and reveals several concerning phenomena. We also propose the design recommendations based on our findings, and discuss the potential of genealogy as a unique lens through which to study a range of important problems in the field of HCI.

\section{Related Work}\label{sec:related}
\subsection{The Genealogy Research Process}
Academic studies of genealogy research practices and community have been scattered across various disciplines, including information science~\cite{yakel2004seeking, fulton2005finding, mansourian2021joyful}, library and archival science~\cite{duff2003list, carini2016information}, sociology~\cite{lambert2002reclaiming, kramer2011mediatizing, bottero2015practising}, tourism~\cite{kennedy2021my, prince2022affect, urrestarazu2022sources}, and public history~\cite{groot2020ancestry, evans2020emerging}. In 2003, \citet{duff2003list} presented a pioneering user study focusing on genealogists' information seeking behavior with archival systems. Their work revealed the complexity of genealogical research, which is more than just ``collecting names and building family trees.'' They found that genealogists need to refer to historical context and apply searching strategies in their research. As archival systems did not meet their needs, genealogists developed their own finding aids, consulted others who had knowledge about a period, geographic area, or community, and formed a strong network through ``courses, genealogical society meetings, and newsletters''~\cite{duff2003list}.

\citet{duff2003list} also pointed out that the genealogical research is an iterative process; experienced genealogists have developed a set of strategies which they can apply each time they look for new information. Similarly, \citet{francis2004genealogy} found a ``circular pattern'' in genealogy research. The information genealogists seek might not be available or easily found; therefore, genealogists often need to circle back to previous steps of their research and look into other research directions. \citet{friday2014learning} summarized these observations and proposed a ``genealogical fact'' frame~\cite{friday2012learning} --- describing genealogical evidence with four elements: name, date, place, event --- to establish an iterative genealogical search model. \citet{darby2013investigating}, at a higher level, suggested that genealogists would start with ancestors who left abundant traces in easily accessible records, then push backward in the timeline to those who are more difficult to research.

\citet{yakel2004seeking} and \citet{fulton2016genealogist} revealed aspects of genealogical research beyond information seeking, which also involves documenting, organizing, producing, and sharing with family members. Intrigued by the story of ancestors, many of \citet{yakel2004seeking}'s interviewees perceived themselves as narrators, archivists, and navigators of their family history. \citet{veale2004discussing}, \citet{fulton2005finding}, \citet{bishop2008grand}, and \citet{moore2021motivates} also elaborated the motivations of genealogists. These studies, through different methods and scopes, characterized genealogists as enthusiasts, explorers and investigators driven by an obligation to preserve family history and motivated by self-exploration through the stories of ancestors. They enjoyed the process of investigation and resolving genealogical challenges, and devoted a large portion of their free time to this hobby.

There are also genealogists who are deemed experts in this field, who have demonstrated their proficiency in genealogy research and provide professional services for individual clients and government agencies. The National Genealogical Society (NGS) in the U.S. recognizes two sources of genealogy credentials: the certification granted by the Board for Certification of Genealogists (BCG) in Washington, D.C., and the accreditation by the International Commission for the Accreditation of Professional Genealogists (ICAPGen) in Orem, Utah~\cite{ngs2018becoming}. Genealogists may earn these credentials by submitting their assessment and documented research project as evidence of their knowledge and research competence~\cite{bcg2021whowhatwhyhow, icapgen2023why}. However, these formal indicators of expertise have not been taken into consideration in previous studies. 
Instead, study participants have been described as ``professional''~\cite{duff2003list}, ``non-professional''~\cite{moore2021motivates}, ``amateur''~\cite{fulton2005finding, fulton2009quid, fulton2016genealogist, moore2021motivates}, ``hobbyist''~\cite{veale2006genealogical, fulton2016genealogist}, or by the number of years they have spent in genealogical research~\cite{yakel2004seeking, yakel2007genealogists, hershkovitz2017genealogy, roued2022search}. As a result, we cannot compare genealogists' research practices except for occasional hints from authors. For instance, \citet{duff2003list} noted that novices might have less knowledge of history background and experienced difficulty converting their need of information to queries of records through archival systems. As other examples, \citet{darby2013investigating} suggested the ``easy tree-build'' stage required ``little expert knowledge'', and \citet{fulton2005finding}'s respondents would not mind hiring professional genealogists for efficiency, but preferred conducting research by themselves for enjoyment. These findings suggest that amateur genealogists might be less efficient in research, but how their research practices might differ from experts' is unknown, and the quality of participants' research work was not assessed.

From an HCI perspective, \citet{willever2014family} were the first to investigate the quality and collaboration of genealogy research on the Internet. They conducted a study of two user-contributed genealogy websites, Ancestry.com and FindaGrave.com (since acquired by Ancestry), and found that the erroneous user production was a major cause of misinformation and conflicts on these websites. Furthermore, participants of their study suggested that Ancestry's advertising strategy and ``research hint'' feature attracted users with little knowledge of the genealogical research process and encouraged them to produce pedigrees rapidly without concern for proper verification. The presence of careless contributors and inaccurate or even fabricated information undermined the overall reliability of user-generated content, introduced burdens to other genealogy researchers, and hindered collaborations on these platforms. However, we still do not have a clear view on how novice genealogists' mistakes are generated in their research, or how technological support may help or hinder the problem.

\subsection{Sensemaking Challenges and Information Literacy}
Genealogical research can be seen as a collaborative sensemaking process~\cite{kim2010tracing}. With a broader definition, sensemaking covers a wide range of human information behavior~\cite{wilson2000human}, spanning information foraging, synthesis, production, and more~\cite{krizan1999intelligence, wilson1999models, pirolli2005sensemaking, zhang2014towards}. As sensemaking pitfalls and challenges have not been explicitly studied in the context of genealogical research, we look for inspiration from other domains.

Intelligence analysis, investigation, and data analytics are common research fields of human sensemaking~\cite{heuer1999psychology, ormerod2008investigative, koesten2021talking}. \citet{heuer1999psychology} summarized how cognitive biases and limitations of memory and perception may lead to analytic pitfalls in intelligence agencies. Studies on criminal investigation pointed out factors that may undermine judgement of investigators, such as social norms on efficiency in the work environment~\cite{ask2007motivational} and asymmetrical skepticism~\cite{marksteiner2011asymmetrical}. Meanwhile, studies of data science found that the lack of clear structure and contextual information prevent users' form understanding and reusing data effectively~\cite{zimmerman2007not, faniel2013challenges, gregory2020lost, koesten2021talking}. These studies often focus on the work of trained professionals, whereas in the genealogist community, a considerable amount of online genealogical work is conducted by amateur hobbyists, who might face more challenges that would not be fully captured in these settings.

Studies of collective sensemaking take into account social context and interactions between individuals~\cite{de2007participatory} and surface more complexities. 
For instance, \citet{kang2014teammate} proposed the concept of ``teammate inaccuracy blindness'' to describe one's unawareness of information quality provided by other team members. \citet{koesten2019collaborative} suggested that when people indeed need to assess data, effective communication with the data producer and other users is beneficial but not well supported. \citet{balakrishnan2010pitfalls} argued that sharing all information between team members at once may enhance initial confirmation biases. Finally, \citet{li2019dropping} analyzed how people made errors depending on the given context at each stage, which then propagated across the sensemaking pipeline.
Another paradigm of crowdsourced sensemaking research focuses on a bottom-up, novice-led environment~\cite{belghith2022compete}, in which the quality of collective sensemaking depends more on individual capabilities, and pitfalls could lead to negative consequences such as misinformation and digilantism~\cite{starbird2014rumors, huang2015connected, del2016spreading, nhan2017digilantism, shahid2022matches}. We have seen deployment and evaluation of crowdsourcing approaches for related tasks such as transcribing and analyzing records~\cite{hansen2013quality, kennard2018computer, wang2018exploring} and identifying historical photos~\cite{mohanty2020photo}, but not one directly applied to genealogy research.

As genealogy research comprises extensive use of archives and libraries, another research field closely related is information literacy, a theoretical framework describing an individual's capability to ``seek, evaluate, use and create information''~\cite{unesco2023information}. This concept was originated from the library and information science discipline. Though traditionally focused on print culture, it evolved in past years --- associated with the concept of digital and media literacy, which emphasize aspects working with digital information and computer --- to address information challenges today~\cite{association2000information, mackey2011reframing, belshaw2012digital}. Information literacy skills were recognized as a critical component in combating online misinformation~\cite{khan2019recognise, sharon2020can, jones2021does}. In the genealogy domain, librarians and archivists focused on educating genealogy beginners frequently encounter audiences with limited information and digital literacy skills~\cite{bremer2018bridging, wilkinson2021digital}. They reflected on amateurs' insufficiency with computer systems and accessing online information, underlining the importance of investigating how amateurs' skills could be improved in a technology-intensive genealogy research context.

\subsection{Learning in Online Communities}

Genealogy was depicted as life-long learning endeavor through which practitioners learn about family history, themselves, and proper genealogy research methods~\cite{hershkovitz2017genealogy}. In this process, amateur genealogists might skip initial learning on how to conduct genealogy research, diving in with a trial-and-error approach~\cite{darby2013investigating}. Genealogical education resources are presented in various forms. \citet{veale2006genealogical} surveyed online educational content for genealogists and categorized it in three groups --- \textit{scholarly}, \textit{topical}, and \textit{ad-hoc} --- arguing that the \textit{scholarly} education was more formal, theoretical, and authoritative, while \textit{ad-hoc} resources had its strengths in personalization, problem-based learning, scale, and immediate availability. \citet{veale2006genealogical} further concluded that genealogists' learning experiences had yet to fully take advantage of a digital, online environment. Participants in \citet{willever2014family}'s study echoed this observation, mentioning that they were willing to help inexperienced genealogists, but the online environment did not provide sufficient support for their educational activities.

Learning is not only the acquisition of abstract knowledge, but also an cultural activity comprising rich social interactions~\cite{national2018people}. Legitimate peripheral participation (LPP), as the core concept of situated learning, describes the process of newcomers to a community starting with participating in peripheral work and gaining mastery along the way~\cite{lave1991situated}. An example that has been extensively studied by HCI researchers is the Wikipedia community~\cite{bryant2005becoming, viegas2007hidden}. However, 
Wikipedia enforces a "no original research" policy, and edits on the site needs to be reviewed and approved by senior members of the community, which essentially constrains community members' use of information. In contrast, in the genealogist community, it appears that anyone has full permission to conduct original research on their family and share research products freely. It complicates the community practice and demands more careful learning experience design for the genealogist community. 

Newcomers are also expected to develop their identity and acceptance of the values and norms of a community of practice (CoP) through participation~\cite{wenger1999communities, handley2006within, cheng2022many}. Whereas various studies suggest that conducting genealogy research enhances people's self-identification~\cite{hatton2019history, shaw2020we, moore2021motivates}, ~\citet{handley2006within} raised concerns that the intention of preserving ``sense of self'' may undermine effectiveness of learning. Furthermore, \citet{roued2022search} found that genealogists often move among multiple genealogical and general-purpose platforms, which may pose more challenges in conveying community norms. Overall, it is unclear from online learning and HCI scholarship how (or whether) amateurs would adopt established practices as members of the genealogist community as they pursue their genealogy research.

\section{Method}\label{sec:method}
\subsection{Recruitment}
To understand the perspectives of genealogists with diverse levels of competence, we recruited participants through two channels: (1) directly inviting professional genealogists via email, and (2) posting recruiting ads on an open online genealogy forum.

For the first channel, we leveraged the member directories of the two major credentialing organizations for genealogical research in the U.S.: BCG and ICAPGen. As of August 2023, the two organizations have 257 and 217 certified/accredited members, respectively, with a few of them obtaining both credentials. We sent invitations with an information sheet of our study to candidates' email addresses obtained from the organization directories. Respondents who expressed their interest in participating were then asked to fill out a survey questionnaire and confirm their intent of participating before we scheduled the interview.

We also recruited from the public mainly for reaching out to genealogists who are relatively new to this field. We posted recruiting information on the r/Genealogy subreddit, the largest genealogy-purposed public online forum we identified with more than 100,000 members. The post contains a brief summary of our study with links to the information sheet and survey questionnaire. We then contacted survey respondents who expressed interest in participating through the email address they provided to schedule the interview. 

In the recruitment survey, genealogists first provided their basic demographic information like gender, age range, and ethnicity. We then asked them about their genealogy research experience, including how long they have been conducting genealogy research, whether they research on their own family, others' family, or for clients as a paid professional. We also asked if they have received formal education/training in genealogical research, if they joined any genealogy related groups, organizations, or online communities, and if they hold any genealogy related credentials. Furthermore, to obtain a preliminary understanding of the research tools and technologies our participants would utilize, we let respondents provide a brief summary of the tools, software and/or websites they would typically use when they were conducting genealogy research. Lastly, we asked respondents to review the attached information sheet and confirm their interest in participating in the interview. The information sheet stated the purpose and procedure of the interview study, participants' compensation for completing the interview (\$20), and that the study is approved by the Institutional Review Board (IRB). A couple of our interviewees generously declined the offered compensation.

\subsection{Interview Design}

We conducted in-depth, semi-structured interviews with participants. As technology-supported genealogy research practices can be multifaceted, our interview
questions were organized into several sections. The introductory questions covered the background research experience of our participants, extending some of the questions in the survey.
The rest were designed based on the high-level research questions described in Section~\ref{sec:introduction}. 
Inspired by literature from this community and academia, we looked into several aspects of genealogy research: (1) How genealogists handle research that lasts a long time without reaching a definitive conclusion; (2) How they overcome research barriers, commonly known as “brick walls”, (3) How they perceive genealogical information and resolve unreliable information, and (4) How they share research findings to others. In each of these aspects, we asked participants how technology has helped or hindered their research. We then asked about their learning experiences. Finally, participants were invited to bring up important aspects about their research that we did not cover. A detailed interview guide is provided in Appendix~\ref{sec:interview}.

\subsection{Data Collection and Analysis}

The recruiting survey responses were collected with Google Forms. Twenty out of 56 respondents participated in our interview study. Half of the interview participants were experienced genealogists who conduct professional research, nine of whom hold genealogy credentials and are recruited from direct invitation, and the other came from public recruiting. This group of participants shared with us their rich experience researching for individual clients, teaching lectures, publishing and editing family history books and journals, working for major genealogy companies, organizations, libraries and government's public repatriation programs, and taking leading positions in local and nationwide genealogical societies. In our later analysis, we refer to them as \textit{experts}. The other half of participants were considerably newer to this field. They had no more than 5 years of experience, did not receive formal education or training, nor did they have professional genealogy research experience. They are referred to as \textit{amateurs} in our following discussion. 

\begin{table*}
  \begin{tabular}{lcccl}
    \toprule
    ID & Gender & Age &\shortstack{Years of\\Experience} & \multicolumn{1}{c}{\colored{Genealogy Education}}\\
    \midrule
    P1E & F & 41--50 & 36--40 & Self-taught\\
    P2E & M & 71--80 & 21--25 & Self-taught\\
    P3E & F & 41--50 & 41--45 & College degree related to genealogy\\
    P4E & F & 18--30 & 6--10 & College degree related to genealogy\\ 
    P5E & F & 61--70 & 11--15 & Week-long lectures, workshops, genealogical society courses\\
    P6E & M & 61--70 & 46--50 & Study group, conferences\\
    P7E & M & 61--70 & 41--45 & Certificate program, genealogical institute programs, study group\\
    P8E & M & 31--40 & 6--10 & College courses, genealogical institute program\\
    P9E & M & 31--40 & 26--30 & College courses, self-taught\\
    P10E & N & 71--80 & 46--50 & Self-taught\\
    P11A & F & 41--50 & 1 & Self-taught\\
    P12A & F & 18--30 & 2 & Self-taught\\
    P13A & F & 18--30 & 5 & Self-taught\\
    P14A & F & 31--40 & 1 & Self-taught\\
    P15A & F & 31--40 & 1 & Self-taught\\
    P16A & M & 31--40 & 3 & Self-taught\\
    P17A & M & 61--70 & 4 & Self-taught\\
    P18A & M & 18--30 & 2 & Self-taught\\
    P19A & M & 41--50 & 2 & Self-taught\\
    P20A & F & 51--60 & 2.5 & Self-taught\\
    \bottomrule
  \end{tabular}
  \caption{Demographic information about our interview participants. The ending letter of each alias indicates the experience level of the participant: A --- amateur, E --- expert. Gender: F --- female, M --- male, N --- not disclosed.}
  \label{tab:participants}
\end{table*}

Overall, our participants are adults between 18 and 80 years old, with a median age range of 41--50. Participants self-reported as White, Hispanic or Latino, Black or African American, and American Indian or Alaska Native ethnicity, with the majority being White. Participants' gender distribution is balanced across the expert and amateur groups. Table~\ref{tab:participants} provides a general description of our participants' research and educational experience. For ease of reading, a letter is appended to each participant's pseudonym as an indication of the his/her expertise, where \textit{E} stands for \textit{expert} and \textit{A} stands for \textit{amateur}.

All our interviews were conducted remotely and recorded with Zoom software. Most of the interviews took about one hour, with a few exceptions extending to 90 minutes. Transcripts of the interviews were auto-generated by Zoom and verified and analyzed manually. The transcripts were anonymized and coded using the Google Workspace package. Specifically, we adopted thematic analysis \cite{clarke2015thematic} to iteratively extract themes from the interview transcripts. These themes were then divided into three high-level topics, associated with either genealogy research practices (Section~\ref{subsec:exhaustive} to Section~\ref{subsec:conclusion}), interactions within the community (Section~\ref{subsec:interactions}), or learning and education in genealogy research (Section~\ref{subsec:learning}). Our coding themes are presented as the headings of sub-subsections in Section~\ref{sec:findings} \colored{ and summarized in Table \ref{tab:findings}}. For each theme, we discuss how amateurs' approach may differ from that of experts and the challenges genealogists are facing during their research, which are emphasized in \textbf{bold} text.

\section{Findings}\label{sec:findings}

Throughout our study, the Genealogical Proof Standard (GPS)~\cite{rose2014genealogical} was frequently referred to by experts to describe their genealogy research process. The Board for Certification of Genealogists (BCG) summarizes GPS with the following five components~\cite{bcg2023ethics} :

\begin{enumerate}
    \item Reasonably exhaustive research;
    \item Complete and accurate source citations;
    \item Thorough analysis and correlation;
    \item Resolution of conflicting evidence;
    \item Soundly written conclusion based on the strongest evidence.
\end{enumerate}

We also found that amateurs occasionally follow similar research practices with experts, applying some aspects of the GPS implicitly without realizing it. Therefore, we adopted GPS as a high-level framework in our later iterations of thematic analysis. We also leverage this framework when discussing genealogists' research process, categorizing our coding themes based on how they contribute to each component of GPS from Section \ref{subsec:exhaustive} to Section \ref{subsec:conclusion}. 

It is important to note that the way we present our findings as experts' versus amateurs' approach does not imply that experts' research is strictly superior. Our participants emphasized that there are hobbyists who do not have credentials but are skilled at genealogy research, and even professional genealogists could make mistakes if they are careless or researching an area that is unfamiliar to them. That being said, we believe it is still important to identify established genealogy research practices realized by expert genealogists, and undesirable ones that are commonly presented in amateurs' research.

Following the findings of genealogists' research practice from GPS, we investigate social context of the community and current learning and educational approaches and constraints, aiming to build a comprehensive view on how amateurs' research and learning can be supported in this community.

\begin{table*}
  \captionsetup{width=.87\textwidth}
  \addtolength{\tabcolsep}{0.3em}
  \colored{\begin{tabular}{p{3.7cm}p{8.2cm}}
    \toprule
    Category & \multicolumn{1}{c}{Coding Themes}\\
    \midrule
    GPS Point 1: Reasonably \newline Exhaustive Research 
    & 
    \textbullet\ Knowledge of records: what’s out there, what information is contained, how to access them\newline
    \textbullet\ Traversing online and local records\newline
    \textbullet\ Focusing on a concise question\newline
    \textbullet\ Strategies for overcoming brick walls\newline
    \textbullet\ DNA as new evidence\\ 
    \hline
    GPS Point 2: Complete and \newline Accurate Source Citations 
    & 
    \textbullet\ Organizing with file systems and family trees\newline 
    \textbullet\ Keeping track with research logs\\ 
    \hline
    GPS Point 3: Thorough \newline Analysis and Correlation
    & 
    \textbullet\ Knowledge of information production\newline
    \textbullet\ Verifying clues by oneself\\
    \hline
    GPS Point 4: Resolution of \newline Conflicting Evidence
    & 
    \textbullet\ Keeping an open mind dealing with uncertainties and new information\newline
    \textbullet\ Resolving conflicts with genealogy timelines\newline
    \textbullet\ Testing hypotheses with family trees\\
    \hline
    GPS Point 5: Soundly \newline Written  Conclusion
    &
    \textbullet\ Producing research reports and family tree data\\
    \hline
    Community Interactions
    &
    \textbullet\ Privacy concerns in online collaboration\newline
    \textbullet\ Peer review and conflicts\newline
    \textbullet\ Perceptions of expertise\\
    \hline
    Genealogy Learning and \newline Education
    &
    \textbullet\ Learning by doing\newline
    \textbullet\ Seeking the right learning resources\newline
    \textbullet\ Knowing without implementing\newline
    \textbullet\ Learning with genealogy software systems\\
    \bottomrule
  \end{tabular}}
  \caption{Summary of findings.}

  \label{tab:findings}
\end{table*}

\subsection{GPS \colored{Point 1}: Reasonably Exhaustive Research} \label{subsec:exhaustive}
Collecting information is the foundation of genealogy research. First and foremost, expert genealogists expect an \textit{reasonably exhaustive} search of information before reaching a conclusion. This requirement has multiple implications for genealogists' research practices.

\subsubsection{Knowledge of records: what's out there, what information is contained, how to access them}

An exhaustive search of information requires genealogists to obtain solid knowledge about the availability of historical documents, including what type of records still exist, what information is contained in each type of records in a certain era, and how to access these records. Expert genealogists demonstrated such knowledge throughout our interviews. As one expert emphasized,
``You have to know where the records are. So that's really the first step'' (P5E).

Such knowledge appears not only helps experts search relevant information exhaustively, but also allows them to obtain an overview of the research scope beforehand and plan their research accordingly:

\begin{quote}
We would generally have an idea of where we want to go and how we wanted to spend the time before we even spent it. We weren't just going around. We knew from the get-go, based on the problem, what record sets we would need to check and how long approximately it would take us to go through a certain amount of records. ---~P8E
\end{quote}

In comparison, amateurs can be \textbf{less knowledgeable about the records}, according to experts, which risks their information search being incomplete or inefficient. Indeed, our amateur participants mentioned their unsuccessful search when they did not know records well enough:

\begin{quote}
There was a fire, and a ton of records were destroyed. Well, I spent a long time trying to find this information, and it literally just didn't exist anymore. And I don't think I realized that until I had already spent a lot of time looking for it. ---~P15A
\end{quote}

\subsubsection{Traversing online and local records}

The emergence of genealogy websites and digitized records in recent decades grants a new level of convenience to genealogists and greatly accelerates their research. One expert described the transformation of this field as a ``massive digital renaissance in genealogy'' (P4E). Expert and amateur genealogists expressed great appreciation for having access to digitized historical documents on the Internet:

\begin{quote}
The computer has made my work so much easier. I can do in an hour what it would have taken me a week to do in the old days. Because I can sit here online even with my crummy Internet, and I can get sources that I would have had to travel hundreds of miles, spent days in a courthouse or in some archives with a pencil and paper doing which I can now do almost instantaneously. ---~P10E
\end{quote}

Along with the record digitization, documents will be transcribed and indexed to support refined searches of data entries with keywords and other metadata. Algorithmic research hints will suggest records and user-built family trees to the users on genealogy websites. Notably, some expert genealogists are skeptical to the inaccurate information that research hints and family trees contain, but they do not object to leveraging these features in their research. As P5A argued, ``There might be a kernel of truth in there.'' Another expert genealogist felt it is necessary to get a sense of the work that has been done by other researchers presented in online family trees:

\begin{quote}
If you haven't done anything and you just want to see what's out there, it makes sense that you would look at the hints that you would look at online trees. You want to see what other people have done, so that you're not recreating everything. ---~P3E
\end{quote}

Genealogists are likely to visit different genealogy websites to access unique record collections and maximize their search coverage. They mentioned even searching through the same record collection on multiple websites to leverage the variation of search algorithms. Still, online public documents is far from the entirety of genealogists' information source. Despite collecting online records, many genealogists suggested that there are reams of local documents not digitized or easily accessible to them, and one has to reach out to many kinds of record keeping facilities like government offices, archives, churches, history and genealogy societies, etc., to acquire these records. Furthermore, expert genealogists suggested that living family members who possess unique private information, such as old family photos, diaries, or even stories, is another viable source of information that genealogists should consult at the beginning of their research: 

\begin{quote}
I want to get a very clear assessment [on] what they know about that ancestor, what they already have gathered about them as far as historical documentation, or what some of the family stories, perhaps, are. And then I like to make sure that there isn't another close family member who may know more than them that we need to potentially interview or connect with. ---~P9E
\end{quote}

Participants complained that there are amateurs who \textbf{stick to a few sources} on genealogy websites, who ``sit at Ancestry,'' ``think all the records are online,'' and have ``limited exposure'' to obscure, local records. Compared with experts, they are likely not be able to satisfy the reasonably exhaustiveness requirement of GPS. Genealogy websites might have a negative impact on this matter, as a few participants (mainly amateurs) suggested that some websites are ``consumer products'' that have financial incentives to monopolize the record collections and keep users within their platform.

\subsubsection{Focusing on a concise question}

Exhaustive search implies the possibility of collecting massive amounts of information for genealogists to process. An ancestor may leave their life traces in all types of documents, and each document contains various information pieces which could lead to new search directions. Therefore, expert genealogists felt that conducting reasonably exhaustive research without a concise research question is impractical. They proposed that genealogists should form concise research questions and try to focus on one question at a time. Furthermore, one participant elaborated that concise questions help genealogists to break down their research into manageable sections:

\begin{quote}
One of the things that that many people don't understand is that, especially when you're doing this as a professional, you really want a very concise question \dots if you don't have a way to say ``I'm done,'' research goes on forever. \dots You really want a tight concise question, and it's better if it's a yes or no. ---~P6E
\end{quote}

On the other hand, multiple amateurs mentioned that they often \textbf{lose focus} in the ocean of documents without a concise research question, and soon be overwhelmed by the information presented to them. Additionally, it appeared that automated research hints might worsen this distraction problem for amateurs:

\begin{quote}
There's something called [a] rabbit hole, where you find a little hint and then all of a sudden your main goal for that day is pushed off to the side and you're chasing a rabbit down this other hole over here to see if it comes up with anything for you. And I find that to be quite the case \dots A lot of documents are getting online and indexed and so sometimes your initial goal for the day is just totally blown away. ---~P20A
\end{quote}

\subsubsection{Strategies for overcoming brick walls}

One important aspect of genealogy research is that genealogists may not always find the direct answer to their question immediately. When hitting a so-called ``brick wall,'' expert genealogists proposed several approaches to overcome it. First, they suggested that re-researching from a different perspective of the subject or looking for leads from relatives around them may help genealogists to find new research leads:

\begin{quote}
Couple of times I've hit the wall \dots and I just put it away for a few days and then start from a different place to try to find it. You know, start from a different family member or a cousin or somewhere else, just completely different than where I was, to see if I can have a breakthrough. ---~P2E
\end{quote}

Second, expert genealogists may advance their approach when there is a big chunk of missing information. Instead of focusing on a particular ancestor, they would research a group of people who share common life experience, such as being members of a family or ethnic community, which suggests shared living patterns among them. Expert genealogists call this approach \textit{cluster research} and suggested it was a more advanced research method not well known to amateurs.

\begin{quote}
There was a gap there of almost a whole generation. You had to sort of use extraneous clues and look for groups of people together.
\dots So you don't want to focus just on the person, the single person. You really need to look for people in groups. Because people moved in groups. They didn't really strike out on their own necessarily. A lot of people ended up really not moving very far. ---~P5E
\end{quote}

A third, more passive approach genealogists brought up was to wait until new evidence surfaces, as there are still many records to be digitized and published. Waiting may sound easy, yet genealogists expressed that they need to deal with negative emotions associated with not being able to solve a problem right away: 

\begin{quote}
It's hard sometimes. You just have to set them aside for a bit until more record collections come out. And sometimes things will appear, and then you can solve it. But sometimes you still never can. \dots It's frustrating, sometimes, though, when you can't solve things you know you should be able to. ---~P1E
\end{quote}

Some amateurs participants admitted that they were driven by curiosity (e.g., ``how far back can I get?''), for fun (e.g. ``playing around''), seeking connections to historical significance (e.g., ``am I gonna find somebody who was at one of the original colonies?''), or aching for an answer to their specific question about ancestor (e.g., ``I was maybe too impatient to try to get the answers.''). For them, the challenge is not to give in to negative feelings accompanied with not achieving their goal, \textbf{rush towards insufficient research, or simply give up}.

\subsubsection{DNA as new evidence}
Genetic testing has emerged in the genealogy community, and these services are now commercialized and integrated into genealogy websites. As people cannot lie about their genetic information, DNA testing is embraced by genealogists as a new type of evidence to complement historical records, which is especially important when there is little other information available, as in researching for orphans, adopted children, or those who born out of marriage. In the meantime, historical records remain the major source for genealogy research. One expert emphasized that historical documents are necessary to explain what is presented in the DNA data:

\begin{quote}
[DNA] is just another piece of evidence. It's not any more valuable than a birth certificate. You can get inferences from DNA, but you can't prove anything with just DNA. \dots You've got to try to find some paper records that give you some insight into why the DNA comes out as it does. ---~P6E
\end{quote}

Genealogists acknowledged that individuals' DNA should be compared with others', and the analysis could become more accurate as more people take DNA tests. As a result, genealogists understand that they might have to wait for more people taking the test to solve their research questions. Some even consider sending DNA test kits ``strategically'' to relatives as a gift and try to persuade them to take the test.

\subsection{GPS \colored{Point 2}: Complete and Accurate Source Citations}

Keeping complete and accurate source citations for research requires good management of information during genealogy research. In practice, we find that genealogists need to document and organize their research and findings along the way for later reference and analysis.

\subsubsection{Organizing with file systems and family trees} 

Organizing research findings was no easy task considering the amount of information genealogists need to collect. Participants mentioned that they store and organize digital records with local file systems and cloud storage. Family trees, either built with local genealogy software or on genealogy websites, naturally present an organized hierarchical structure, and genealogists can attach documents, citations, and add notes to ancestors' profiles in the tree.

\begin{quote}
During research, the Ancestry tree was actually a pretty active part of the process. It was where we 
\dots stored all the information relating to people and families and stuff. If there was a document that we found if it was on Ancestry, we would save the Ancestry version to the Ancestry tree. If we found it elsewhere, we would either download the image and upload it, or we would scan it and upload it to the tree. ---~P8E
\end{quote}

Organizing documents and citations can be tedious and time consuming. Many amateurs struggle at this task, as they often described their research being ``messy,'' and one expert commented on amateurs that ``half the battle is because they're not organized'' (P1E). We observed that amateurs may treat organization as a secondary task, and \textbf{make little effort in organizing} their research. As one amateur noted:

\begin{quote}
It's just like a really ugly draft of my research. And I don't want to put a lot of effort into making it pretty and ``putting the bells and whistles'' on it just to find out it's wrong, or I'm missing something, or I gotta kind of redo it. ---~P14A
\end{quote}

However, as we will demonstrate in the following themes, experts also face the challenge regarding the uncertainties of their pending conclusions in their research, and they develop a set of tools to help accommodate it.

\subsubsection{Keeping track with research logs}

\begin{figure}[h]
  \captionsetup{width=.87\textwidth}
  \centering
  \includegraphics[width=0.9\linewidth]{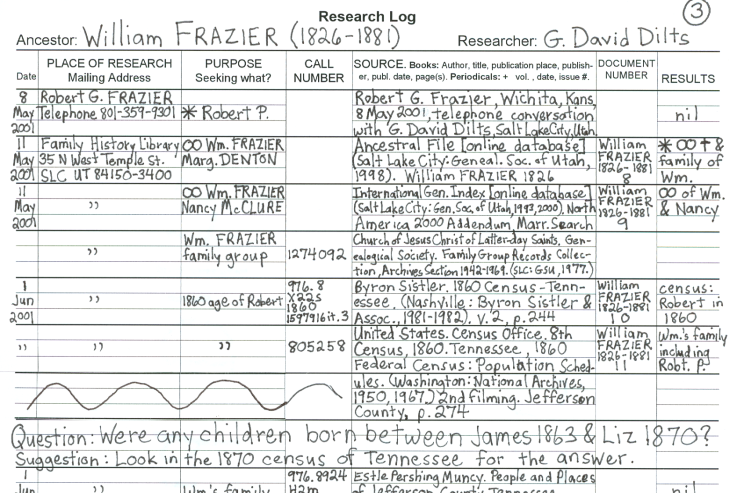}
  \caption{An example of paper-based genealogy research log from \cite{familysearchwiki2023log}. Note that it contains the name of the ancestor being researched, a research question, and detailed documentation of means, purpose, and result of the search.}
  \label{fig:log}
\end{figure}

Many expert genealogists asserted that not only findings are worth documenting. In fact, they developed \textit{research logs} (see an example in Figure~\ref{fig:log}), which are spreadsheets that document genealogists' searching activities to help them keep track of research progress,  even if they end up with no significant findings. The benefit is multi-fold according to experts: First, the logs record information about their search attempts, such as what collection is searched, what keywords were used, in which websites, etc., so that they could avoid duplicating the search effort when picking it up later. Second, a research log states the current research question for this research session to help genealogists stay focused. Third, research logs help genealogists store information they found that is not directly related to the current question but might be useful as the research progresses, so that they can refer to it when necessary. Finally, for professionals, a research log helps them demonstrate to the client what they have spent time for during the research session. This is important as most professional genealogists charge based on the time they spent on the client's project. One expert shared with us a very detailed use of research logs:

\begin{quote}
A lot of genealogists, professionally, use some kind of a log to track their searches. \dots And I just keep track of what I looked at, what source I consulted, and then I write down what was my purpose and intention for the search, what was I looking for. And then I also track what were the specific criteria that I used. So if it's a database, ``Did I search for all people with just the first name?'' ``Did I look for all people with the last name?'' ``Did I use a wildcard to give me spelling variation?'' So I'll keep track of the search parameters, just so that I know how thorough my search was, and then I'll make notes about the outcome. ``Did I find some potential candidates?'' ``Did I find the ancestor I wanted?'' Or maybe I didn't. So I keep track of all of that, and I make comments. ---~P9E
\end{quote}

\textbf{None of our amateur participants mentioned keeping research logs} like experts did. They often seemed unaware of the benefit it may bring to them. Amateurs proposed alternative approaches to help them keep track of the research progress, such as appending notes and tags to family trees, writing down paper notes, and keeping browser tabs open. Unsurprisingly, they seemed to struggle at times and found this approach ineffective:

\begin{quote}
I think to myself, ``Oh, this is great. I need to remember to put this over for this person'' and then. You get distracted, you have to go do something else, something comes up, then you come back the next day and you go, ``What was it I was thinking about?'' So I don't currently keep a log or a record of what I'm doing and I should.
---~P19A
\end{quote}

\begin{quote}
I'm the person that'll have 30 tabs open, and I'll be opening tabs until it forces me to stop, or, you know, I've got a little notebook next to me that I'm making weird gibberish notes. And then if I go back a week later, I'm like ``I don't know what the hell this means!'' ---~P15A
\end{quote}

\subsection{GPS \colored{Point 3}: Thorough Analysis and Correlation}

The third point of GPS suggested that genealogists need to evaluate (through analysis) and summarize (i.e., correlate) all the information collected that is relevant to the research question to draw a conclusion. Expert genealogists take cautious steps in this process, noting that research of each ancestor is built upon that of closer generations in the family tree. Genealogy literature introduced metrics in evaluating the \textit{source}, \textit{information}, and \textit{evidence}, each of which has a specific meaning in experts' genealogy research. As a primer, sources are records in various forms which could be \textit{original} if the record is not derived from other sources, \textit{derivative} if it is, or \textit{authored} if the author's opinion and observation is involved. A piece of information is considered \textit{primary} if the informant has firsthand knowledge of the described event, \textit{secondary} if they does not, or \textit{indeterminable}. Lastly, evidence is the body of records that are relevant to the research question, which can be categorized as \textit{direct}, \textit{indirect}, or \textit{negative} based on whether the evidence could directly answer the question~\cite{rose2014genealogical}.

\subsubsection{Knowledge of information production}

Expert genealogists demonstrated awareness of how information was produced and how errors and mistakes could be made in the process throughout our interview study. They described in detail the cultural background of historical records and why the information might not be reliable in some of them, the process of original documents being digitized, transcribed and indexed for easy access and search, and how record hints are generated from indexed records and others' trees on genealogy websites. 
Genealogists essentially use such knowledge to assess the quality of information they gathered. The following examples shows how deeply genealogists understand the information they are working on. For instance, on original documents:

\begin{quote}
``\dots Well, it's in the census record. It has to be right!'' And I said, ``Look at that low X shows who gave it. It was an 8-year-old daughter. `Tell me when, [in] what year, your parents were born in Ireland!'\thinspace'' ---~P2E
\end{quote}

On record hints:

\begin{quote}
``As far as record hints go, that is only as good as the computer algorithms and what is already in the tree. \dots
So it's only searching through the small amount of indexed information, and matching it to what people have put in the tree, whether it's right or wrong.'' ---~P3E
\end{quote}

Amateur participants did not describe the ins and outs of the information production procedure as precisely as experts. It appeared that amateurs \textbf{sense the reliability of documents from experience}, who have also expressed their uncertainly in assessing some of the records they encountered:

\begin{quote}
I have found they [church records] are very thorough in terms of noting the names of everybody involved, birth dates, christening dates and so on, marriage dates. They're really pretty thorough at that time. So I do think they're accurate. But then there's other documents, like in 1920 [census] in the United States, that I'm not quite sure are as accurate. ---~P20A
\end{quote}

\subsubsection{Verifying clues by oneself}

Even though expert genealogists are willing to leverage research hints and others' trees in their research, they are also selective when evaluating if one piece of information can serve as evidence in formulating their conclusion. In general, experts expressed their preference of using primary information from original records as evidence, while considering the rest, such as others' family trees and algorithmic research hints, as research clues that they need to verify themselves. In P6E's words, 
``I'll take it [someone's tree] as a hint. And I’ll do what I have to do to validate it.''%

Besides re-approving everything in others' family trees, genealogists also develop strategies to effectively assess the overall quality of a family tree. In particular, they mentioned several defining characteristics of a tree, such as sparse supporting evidence, conflicting or duplicate documents in a profile, and overly large or deep family trees, indicating that the tree might be poorly researched or copied from somewhere else. For instance, one participant described using the ratio of records to ancestors in the tree in his evaluation metrics:

\begin{quote}
There's no way most people have enough records, enough proof of who 6,000 people are to put them in there. \dots it'll show, say, 6,000 people, 6,500 records. So you go, ``You have barely over one record per person. 'Cause I have people with 20's and 30’s [records per person], that tells me that you have not clearly looked at their stuff.'' ---~P19A
\end{quote}

Additionally, experts asserted that amateur genealogists who are less competent often \textbf{treat clues as evidence without proper verification}. Several amateur participants of our study admitted the problem of they accepting research hints to build family trees without any verification at the beginning of their research. P16A reflected on his initial research on genealogy websites:

\begin{quote}
I did initially, but I no longer sort of blindly accept the connections that other users on Ancestry did. Like the first weekend that I had my Ancestry account, \dots you have like a little shaky leaf that creates hints, and I thought that I had successfully traced my ancestry back to like 1200 [CE] England. ---~P16A
\end{quote}

Participants further acknowledged that such behavior is common among amateurs. As P18A elaborated, ``There's a big problem in genealogy community at most of the beginner's stage \dots where I'll just copy information over and over.'' Considering how easy for users to accept research hints and duplicate others' family trees on genealogy websites, one can quickly expand his/her family tree by replicating unverified information. However, expert genealogists emphasized that the frequency of one piece of information does not imply its accuracy. P1E argued, "Just because it's there 40 times doesn't mean it's true."

\subsection{GPS \colored{Point 4}: Resolution of Conflicting Evidence}

Even when working with original documents, genealogists will likely find conflicting information among the records. Participants mentioned two types of conflicts. First, they suggested that various records may not belong to the same research subject, even though partial information (name, age, location, etc.) matches. Second, there are also cases where conflicting information exists due to the inaccuracy of historical documents, as discussed in the previous theme. Therefore, it relies on genealogists' expertise to determine if all the information they collect is accurate and actually belongs to the same person. Furthermore, circling back to the first point of GPS, only after reasonably exhaustive research can genealogists confidently determine that all conflicting information is identified and resolved before reaching the final conclusion.

\subsubsection{Keeping an open mind dealing with uncertainties and new information}

First and foremost, experts acknowledged that following GPS does not guarantee perfect conclusions, as new records might appear at later time that overturn a conclusion:

\begin{quote}
The genealogical proof standard doesn't guarantee that your answer is correct. But it makes it more likely that it is correct. It's more likely that if new information surfaces that your conclusion will stand. ---~P7E
\end{quote}

As a result, expert genealogists try to keep their mind open to other interpretations and new pieces of information. One participant stated:

\begin{quote}
A lot of genealogists don't ever say they've proven something. They say ``the evidence shows,'' or ``it looks like this is the answer'' or ``this is probably the correct thing.'' And we're always open to more interpretation in the future as more records become available. ---~P3E
\end{quote}

Additionally, several participants reported that they often met amateurs who \textbf{hold predetermined beliefs} from family stories and their own research, which prevented them from accepting evidence that indicated otherwise:

\begin{quote}
When people hear a family story sometimes they have a really hard time letting go. \dots If people get so fixated on, you know, these things that they've heard, and they've believed to be true, that sometimes, even when you walk them through how records show that it would be impossible, they just don't believe you. ---~P4E
\end{quote}

\subsubsection{Resolving conflicts with genealogy timelines}

\begin{figure}[h]
  \captionsetup{width=.9\textwidth}
  \centering
  \includegraphics[width=.9\linewidth]{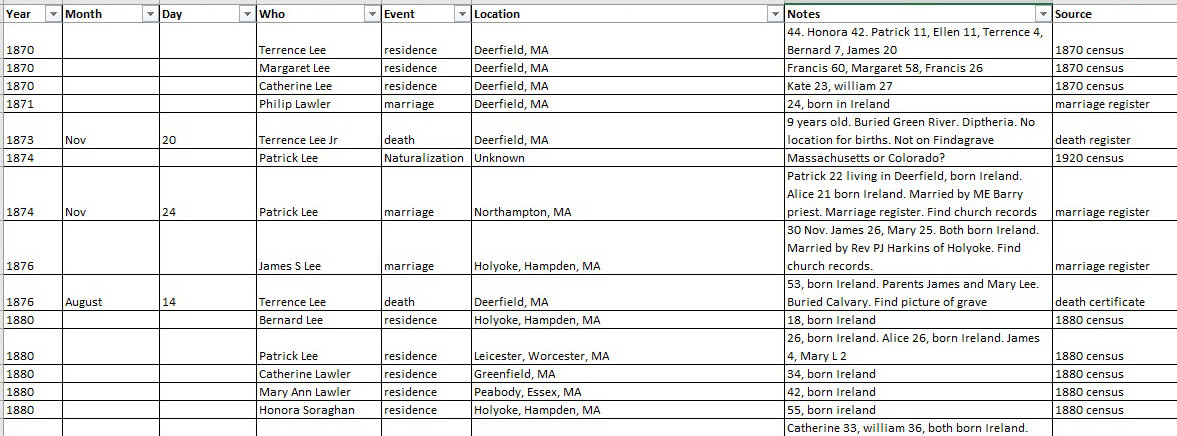}
  \caption{An example of genealogy timeline from \cite{neaves2021simple}. It lays out chronological and geographical information of multiple ancestors to help a genealogist to identify research gaps and conflicts.}
  \label{fig:timeline}
\end{figure}

In practice, genealogists developed the timeline tool, which is a spreadsheet that lays out \colored{ancestors'} lifetime activities in sequence based on the collection of records (see Figure~\ref{fig:timeline}). By presenting information in this way, expert genealogists can spot conflicts by identifying events that are impossible to occur given natural constraints and historical context, such as children being born before their parents or people moving between distant places within a very short period of time. Furthermore, the timeline tool can help genealogists to recognize missing information in certain periods of an ancestor's life. As one expert summarized:
\begin{quote}
Timelines are another really important tool that we use. 
It's particularly [useful] if you have two people with the same name when you're trying to separate them out and figure out which John Smith is this one. \dots If you lay out a timeline, and where they are, and what the date is, you can sort of figure out which of these facts goes with this particular person. \dots It also helps you figure out where your holes are. ---~P5E
\end{quote}

Similar to research logs, amateurs \textbf{didn't specify creating timeline sheets themselves}. We notice that some genealogy websites provide timeline functionality for their users. However, these timeline features tend to be constrained to an individual ancestor's level, which is not as flexible as experts' customized spreadsheets, and may hinder genealogists' analysis across multiple research subjects.

\subsubsection{Testing hypotheses with family trees}

When exploring alternative explanations, several genealogists also mentioned that they would build hypothetical (or speculative) family trees to help them test their research hypotheses. We heard different approaches from our participants on this matter across our interviews. One participant suggested that she would use the ancestor profiles to store unexplored leads:

\begin{quote}
Once I get the first document that indicates, ``Okay, they might be related to me through this other person,'' then I will create a profile for them, basically just adding their name to the tree and who they're connected with, but there really isn't much on it. I don't add in any facts until I can find the documentation for them. So I just mainly create the little profile just to keep track of their name. But no, I don't really rely on it until I can find more information. ---~P20A
\end{quote}

Another participant mentioned that she would build the ancestor profiles first by accepting hints, and resolve conflicts in it afterwards:

\begin{quote}
I would kind of do a rough draft, accepting hints, and then kind of go back through it and check my work and add legitimate records and census records. If I couldn't find anything then I would delete it. But, usually, \dots if you come back a couple of days later, there's gonna be more hints for records, and then you can kind of cross reference your records, and make sure that everything fits. ---~P15A
\end{quote}

We note that \textbf{hypothetical work is often part of users' public family trees}. In such cases, other genealogists may have to refer to notes and tags in the tree, or rely on their own interpretation to determine which part of the tree is thoroughly researched and which is not. Some genealogy websites allow building multiple trees and keeping them private, so users have an option to detach hypothetical work from the public (i.e., proven) family tree. And one expert mentioned working with two genealogy software to separate proofs and hypotheses.

\subsection{GPS \colored{Point 5}: Soundly Written Conclusion} \label{subsec:conclusion}

The final step of genealogists' research is to express their research conclusions in appropriate formats. Expert genealogists treat their research product seriously, tying the quality of it to their professional reputation.

\subsubsection{Producing research reports and family tree data}

Research reports are a very common product in expert genealogists' work. The format of a report may vary based on clients' needs, but generally, it will contain all the (references to) documents genealogists have collected that are related to the research question and a thorough explanation --- expert genealogists called it proof argument --- on how they reached the conclusion based on the evidence. Expert genealogists may publish their research (sometimes on behalf of clients) in genealogical journals or as books. They use research reports to showcase how they follow the GPS throughout the research, and others can also use it to evaluate the quality of work:

\begin{quote}
You have to give people enough information, preferably, so that they could go and find it [evidence] themselves and confirm that you actually interpreted it correctly \dots You have to give people enough information so that they can judge the quality of your work. If you see a genealogical work product without documentation, as is often said, that could be fiction somebody have just made up. ---~P7E
\end{quote}

In practice, family trees and associated data sets can be an important complement to genealogists' research deliverables. Genealogists have developed open-source data formats, such as GEDCOM (an acronym for Genealogical Data Communication)~\cite{harviainen2018genealogy, wikipedia2023gedcom}, that are widely accepted by the community for storing and transferring genealogical data. In addition, an expert noted that online family trees have made it convenient for them to keep clients posted about the research progress.

We found that many amateurs are \textbf{enthusiastic in sharing their findings online, but not necessarily in formal reports}. Their sharing is motivated by the intent of reaching out to relatives and altruism to descendants and other genealogists, as one participant mentioned:

\begin{quote}
I really wanna share what I find. I know it's maybe a small group of people who might be interested in that, or someone might be interested in it years from now. \dots Because it's no harder for me to do that than it would be to keep my own records and then it has the added benefit of maybe helping someone else now or down the road. And certainly I've appreciated that when people have been transparent and had public trees and things like that on Ancestry or on FamilySearch. ---~P16A
\end{quote}

Similar to organizing, experts agreed that producing a report with full industry-standard citations is ``very time-consuming'', and some of them may leverage genealogy software to automate part of the report generation process. Amateurs, on the other hand, simplify the process by self-publishing family stories or appending records and notes to their family trees to express their conclusion. Meanwhile, our findings suggest that there might not be a community norm on how a family tree should be built and presented on the Internet. It could be a combination of verified work, hypotheses, and research leads, depending on individual genealogist's preference.

\subsection{Community Interactions} \label{subsec:interactions}

Social context is important when investigating the learning experience of community members. In genealogy specifically, as one participant commented, ``there is a huge social component,'' noting that the self-identity aspect of genealogical research is ``super tender to people'' (P11A). In this section, we focus on social interactions in this community, rooted in the practices of genealogy research that surfaced throughout the interview study.

\subsubsection{Privacy concerns in online collaboration}

It is common for genealogists to seek help from their peers in online groups. Our participants shared with us several reasons: specialized or technical assistance such as analyzing complex records or DNA data, translating documents from a foreign language, and restoring aged photos; as well as seeking research advice more generally. Many of these services (except for image restoration) are provided by community members free of charge. Similar to genealogy websites, amateurs may decide to jump across different online groups based on their specific information needs. Genealogists expressed great appreciation of these online groups for their altruistic work:

\begin{quote}
I'll send my students there [a Facebook group] because it's really really helpful. If I have a question, people can answer it, and there's experts there that will answer and they do it very quickly. So it's a great way to get help, and with little cost or no cost. ---~P1E
\end{quote}

We heard about diverse practices for sharing genealogy information from our participants. For some amateurs, their enthusiasm in sharing genealogical information goes beyond genealogy-focused websites, extending to general purpose websites like social media. Several participants expressed concern for the \textbf{risk of revealing personal information} in genealogy research, who may limit their appearance on social media and provide just enough information to an online community when asking for help. It may make the online collaboration less effective, but their privacy concerns are not baseless. Government records are usually released to the public decades after they are generated for privacy reasons~\cite{heimlich201272}. Genealogy websites follows a similar rule, limiting access to profiles on living people in the tree \cite{ancestry2023living, familysearch2023family}. In the meantime, more than one amateur participant mentioned researching living individuals who they think might be their relatives using public online information:

\begin{quote}
I then, to be honest, I then get online, get on Google. If they have any information at all, you know, if they have a unique name --- I've also tested at 23andMe and so if they're on 23andMe then there's usually a little bit more information --- sometimes there's their age, sometimes it's what city they live in or what state they live in and I can just start Googling their name and a potential age and start to identify them. \dots I'm basically, if you are familiar with the term, ``doxing'' someone. ---~P17A
\end{quote}

\subsubsection{Peer review and conflicts}
Some expert genealogists emphasized the importance of sharing and peer-review, in a scholarly or publishing context, as a means to ensure genealogy research quality. Some experts expected this approach to be applicable on genealogy websites as well, while one expert acknowledged that the growing diversity of sharing mechanisms makes community moderation challenging:

\begin{quote}
The way that genealogical information is shared has really moved towards these forums, websites, or self-published materials. You're not really seeing it mediated quite so much by genealogical society. ---~P8E
\end{quote}

In practice, many genealogists expressed \textbf{great difficulty in correcting others' (potential) mistakes} on the Internet. First, it is almost impossible to correct inaccurate information propagated across data sets and individuals' family trees online. One expert shared us an example of himself:

\begin{quote}
That piece of information has been copied through GEDCOM and GEDCOM and GEDCOM, and copied up into Ancestry file after Ancestry file. Everyone's pulled it up on FamilySearch. \dots They've copied it at MyHeritage. It's just out there and there is no way to pull that back. \dots It's kind of like standing in front of a landslide and going, ``Stop! Stop!'' It isn't gonna happen. ---~P5E
\end{quote}

On the other hand, recipient of critics and correction suggestions could feel offended and ignore the message. For instance, a participant expressed frustration when sharing her most recent findings on r/Genealogy subreddit, for other people suspected there might be mistakes in the conclusion:

\begin{quote}
I just don't know if I wanna go back to justify to that group how I know what I know. \dots Some of the posts about it were not very polite \dots I'm happy to maybe answer questions if you didn't try to insult me about how I found this so quickly. \dots Even as helpful as those people are on other things, every once in a while they \dots 
just get a little rude, a little toxic. ---~P20A
\end{quote}

Last but not least, many genealogists who found potential mistakes often feel hesitant to contact the author, in fear of hurting others' feelings, messing up another user's tree, or getting involved in an argument, regardless whether they know the author personally:

\begin{quote}
Within the past weekend I found a huge error on one of my cousin's trees, and I haven't been able to tell him yet. I don't know how, and I know he's spent years and years on this research, but he got one of the people wrong in the very beginning. \dots And with this particular cousin I have editorial status that he's given me on his tree. So I can actually go in and fix it, but I don't wanna mess up all of this work. ---~P11A
\end{quote}

\begin{quote}
I'm not confrontational, so I'm not gonna call anyone out. \dots On one hand, I probably take it more seriously than I should, because I just get obsessed with [genealogy]. But, on the other hand, I don't care enough to get into an Internet fight with anybody about something. ---~P15A
\end{quote}

\subsubsection{Perceptions of expertise}

Expert genealogists are not free from the conflicts described in the previous theme, even when working with or for amateurs. Experts shared with us their experience when clients were not satisfied or disagreed with their research outcome. They did not find solutions to resolve the conflict, except for acknowledging different opinions or being more selective taking clients. In one case, P4E recalled, ``My client didn't believe me, and so we very graciously and politely decided to part ways.'' P9E reflected, 
``Through my years of experience, I'm a lot better at selecting the clients that I wanna work with, and that I know will be good partners and working with me.''

Similarly, genealogists' expertise does not grant them special authority in most online environments. They understand that the family trees they build on open genealogy sites are modifiable by anybody else, regardless of their research capability. One expert also stressed the challenge of content moderation on genealogy websites, and hoped that effective education of the amateurs might help mitigate the problem: ``There's not a really good way to say `you're banned from posting about genealogy for the rest of your life,' but I feel like the more people I teach how to do good research the better the research will be eventually'' (P3E).

A few experts, however, intended to \textbf{disengage from intensive online interactions} with the public --- often after an initial attempt of sharing their knowledge --- to avoid the wariness and complications. As one participant summarized:

\begin{quote}
There are genuinely kind people that will donate their time and share this advice freely. And there are plenty of people like that. But it seems to me [that] people with the most experience tend to veer towards really just providing their analysis while they're on the clock, \dots or operating as a specific capacity for their job \dots I think a lot of us professional genealogists all just kind of go through this initial phase where we join every group possible and want to share everything we've learned, and then we get burned out. And then we just focus on our work in our own little corner. ---~P8E
\end{quote}

We also asked participants about financial relationships between genealogists. We found that not every amateur would like to hire experts to research for them. Cost is certainly one factor of concern. More importantly, they expressed \textbf{preferences of researching by themselves} to enjoy the experience of solving mysteries, gaining knowledge, and exploring historical context that connects to ancestors' life through genealogical research:

\begin{quote}
The fun of it is discovering it for myself instead of just having someone hand it over for me. It's just being able to feel connected and immersed in, and then feeling good and enthusiastic about the connections or things that I've been able to find or validate ... So I wouldn't be opposed to it [hiring professionals]. I think that they bring a ton of value, but I definitely think it takes away a little bit from the experience. ---~P14A
\end{quote}

Those who consider hiring experts may do so for acquiring local documents, overcoming brick walls or double-checking their own research. Amateurs don't blindly accept experts solely based on credentials. Instead, they \textbf{assess experts' research capability} using various criteria, such as how physically close experts are to the record collection, the professionalism of service's website, and affiliations with prominent organizations.

\subsection{Genealogy Learning and Education} \label{subsec:learning}

As we have discussed, amateur genealogists often conduct original research on family history, and individuals' research ability has great impact on the quality of the overall product on genealogy websites. Therefore, understanding how amateurs learn to conduct good genealogy research in a technology-mediated context is critical for this community. 

\subsubsection{Learning by doing}

Learning-by-doing is a common approach mentioned by both amateurs and experts, many of the latter having started genealogy research as a hobby. Amateurs described ``jumping into'' genealogy without much knowledge about proper research methods, relying on what the software or websites provided them. In P15A's words, ``Honestly, I guess Ancestry.com itself is \dots just user-friendly enough that I figured it out.''

Amateurs admitted that they could make \textbf{lots of mistakes at the early phase of research} with this trial-and-error approach. Their learning is triggered when realizing previous mistakes or encountering research barriers. As their knowledge grows, amateurs described researching more rigorously to review, improve, or completely redo previous work. However, we also found that some of the early mistakes could remain in the family tree. It was unclear whether every amateur would recognize their mistakes and start learning the proper research method soon after they started genealogy research.

\subsubsection{Seeking the right learning resources}
Genealogists may need to learn not only research methods, but also facets of many other subjects, such as foreign languages and history, to advance their research. Amateurs' learning is typically self-directed. They need to \textbf{proactively look for content that is appropriate to their knowledge level and information needs}. For example, P13A said,
``I don't know where my German ancestors came from, so I need to learn about the history of Germany, so I can narrow it down by what region they might have come from.'' %

Amateur participants mentioned free online resources for learning genealogy research, like YouTube videos, podcasts, and blogs, with video lectures emerging as among the most common sources for learning. Even if amateurs did not actively participate in online groups, they considered browsing others' posts an important and effective method to learn from them:

\begin{quote}
I look for posts \dots and read the follow-on posts for further ideas about how I can do the work that I'm interested in doing. I look at it, I monitor it and I read the comments looking for more ways to to expand my research to do something better. --- P17A
\end{quote}

Experts endorsed many of the free online resources that amateurs mentioned. Furthermore, a few of them proposed that the FamilySearch Research Wiki~\cite{familysearchwiki2023familysearch}, a web-based encyclopedia of genealogy research, could be a good source for learning, but it was rarely brought up by amateurs when they described their learning. One amateur, P13A, even explicitly requested a ``Wikipedia style'' site apparently without the knowledge of FamilySearch Research Wiki, which may indicate a disconnection between learners and learning resources.

Experts also suggested more formal resources for newcomers to advance their research, including seminars offered by genealogical societies, degree and certificate programs from higher educational institutions, genealogy journals, and professional study groups.
Depending on amateurs' allowance and research progress, formal education might become more of an option when their knowledge on and dedication to genealogy increases. P20A explained her evolving priorities:

\begin{quote}
I think I'm probably reaching the point where I would go to seminars and things. I was maybe too impatient to try to get the answers that I wanted at the beginning. And now that I've got a lot of those answers. And the urgency isn't there quite so much now, I might be willing to go and just learn about the technique of genealogy research.
\end{quote}

\subsubsection{Knowing without implementing}
We also heard from amateurs that they may \textbf{not apply their knowledge on genealogy research in practice}, which suggests that having knowledge on genealogy research methods does not directly translate to their implementation. As P16A mentioned, ``I've listened to some genealogy podcasts and I've seen some webinars. I'm familiar with the formal research process, \dots I have never gone about researching genealogy in such a formal way, like by having a literal question that's written down with supporting evidence.''

One reason we identified is that amateurs may under-appreciate the benefits of established research methods due to less experience and a misalignment of research goal and method. For example, one participant spoke to the tension between her primary motivation to tell a good story about the family, and the suggested method of explicit and structured organizing and analyzing:

\begin{quote}
There's so much out there and I've watched a lot of YouTube videos and tutorials on the tried and true and the best practices and how to organize and lay it out. I struggled with that method a lot in the beginning because it wasn't my vision. A lot of it \dots have these very structured trees, and it's just the facts, and that just goes down the lineage. But to me, I didn't connect with that. I wanted my book to tell a story about their lives and their experiences, and how it related to events, and really be able to connect. ---~P14A
\end{quote}

\subsubsection{Learning with genealogy \colored{software} systems}
A couple of expert genealogists asserted that amateurs can accomplish much more nowadays. Optimistically, they suggested that modern amateurs' skills can sometimes be on par with experts of previous eras, thanks to emerging technologies in the genealogy industry:

\begin{quote}
So many people now can get started and make a lot more progress that 20 years ago, I know professionals who were working 20 years ago that aren't working now, because the clients have matched or outpaced their skills. \dots The client can now come to us with so much more prep work than they used to, because now they know how to, where to go, and they can get so much more very quickly. ---~P9E
\end{quote}

In contrast, amateurs themselves asserted that \textbf{genealogy \colored{software} systems could have been more helpful} in educating newcomers about genealogy research.

\begin{quote}
I wish Ancestry did a better job of explaining basic methods as you're starting a tree \dots They haven't educated the consumers on what to do before starting this. \dots They're just like, ``Here, go at it!'' without kind of explaining the methodologies, and why you cross reference all these different things to make sure you have the same person. ---~P11A
\end{quote}

\begin{quote}
I got DNA tested on ancestry. I got DNA tested on 23andMe. And I don't believe either one of them had a good ``Here's how to get started.'' \dots 
I don't believe either of those sites have that, from a beginner point of view, ``what to do'' and so forth. ---~P17A
\end{quote}

We asked participants what functionalities they would like to see to help with amateurs' research. From a high-level view, experts believe that amateurs would benefit from ``hand-holding'' instructions with prompt question-answering during genealogy research:

\begin{quote}
A lot of beginners are just like, ``Well, I just want a step-by-step thing, maybe like with the video, or just where I can stop and ask questions.'' And I think that's something where technology would play a huge role in automating that and then updating that. ---~P4E
\end{quote}

Amateurs provided more detailed requests. They imagined that AI-powered \colored{computer} systems could explain proven research approaches to users, propose research directions, introduce new sources, and present historical context during the research: 

\begin{quote}
Tying it to historical events definitely brings a lot of value. And so if you're pulling these census or pulling these certificates, or you know these immigration record or whatever, \dots the AI would be ``hey, these are also recent historical events.'' And I know Ancestry does have kind of a similar timeline, but I don't think it's integrated when you're searching. --- P14A
\end{quote}

\begin{quote}
[If] I buy an old tax record, I can't be sure if it's accurate, right? Maybe AI could help me determine if it is accurate, and then beyond that could tell me what next I could do with it, and give me 3 reasons why that's relevant to my life today 150 years later. ---~P16A
\end{quote}

Automatically verification of records and timeline is another desired features among genealogists. They suggested that \colored{computer} systems could flag questionable conclusions and conflicting evidence to help amateurs improve their research quality and assess the quality of other's work more effectively. In P10E's words, ``They could have a verifiable system where you would have some reliability, and that would also help the amateur, because it would force the amateur to go do the research.''

\section{Discussion}\label{sec:discussion}

\citet{wenger2011communities} described Communities of Practice as ``formed by people who engage in a process of collective learning in a shared domain of human endeavor.'' As genealogy research practitioners build family histories with available information from various types of historical documents, we describe the genealogist community as a \textit{community of information practice}, in which genealogists learn best practices involving information foraging, analysis and production. Our study revealed that genealogy research practices resonate with frameworks of sensemaking and information practice proposed by prior research in other domains. For example, \citet{krizan1999intelligence} proposed an intelligence analysis process that comprises the conversion of information needs into intelligence requirements, in addition to information collection, processing, and production. \citet{pirolli2005sensemaking}, in their influential work on sensemaking, also described a process of foraging information, extracting evidences, developing structured schema and hypotheses, and constructing final presentation. From an information literacy perspective, the Association of College and Research Libraries (ACRL) also stressed the importance of understanding information creators’ expertise and credibility, the necessity of managing information, and awareness on ethical and legal use of information~\cite{association2000information, association2016framework}. All of these elements were reflected in genealogists' research routines in our study. 

Yet, the challenges our participants experience are not exclusive to the genealogy domain. Amateur genealogists' approach to accepting hints echoes \citet{kang2014teammate}'s observations of ``teammate inaccuracy blindness,'' which in turn leads to the propagation of inaccurate information in an asynchronous collective sensemaking process. \citet{li2019dropping} also discussed an increasing number of errors in their crowdsourced sensemaking pipeline due to ``insufficient context and missing information'' from early steps. Their findings on different types of errors, such as inattention to background knowledge and low effort errors, also find their counterparts in amateurs' genealogy research behaviors. As amateur genealogists struggle with assessing quality of historical documents due to their insufficient knowledge, \citet{koesten2019collaborative} similarly stressed the importance of data creators communicating the original purpose and limitations of data with its users. Our study contributes evidence towards the universality of these collective sensemaking challenges at an online community scale.

Importantly, experts in the genealogist community developed research standards heuristically, attempting to ensure quality of the community’s research outcomes by regulating essential aspects of information use and production for individual genealogists. Experts advocate for explicit distinction among different types of information sources (original documents, research hints, others' family trees, etc.), critical evaluation of information (based on its quality, correlation with other information, and relevance to the research question), and thorough documentation of research process and reasoning. Their standards and practices provide useful guidelines for us to identify challenges in amateurs' approaches and ways to support them in genealogy research. We summarized these challenges in Table~\ref{tab:summary} and explore mitigating approaches.

\begin{table*}
  \captionsetup{width=.87\textwidth}
  \addtolength{\tabcolsep}{0.3em}
  \begin{tabular}{p{1.6cm}p{10cm}}
    \toprule
    Category & \multicolumn{1}{c}{Challenges of Amateur Genealogists}\\
    \midrule
    Inadequate \newline research & 
    \textbullet\ Unfamiliarity with records and research methods \newline
    \textbullet\ Broad, shifting research questions \newline
    \textbullet\ Scarce, unorganized documentation of research process and conclusion \newline
    \textbullet\ Insufficient analysis and correlation of evidence \newline
    \textbullet\ Impatience and biases \newline 
    \textbullet\ High-productivity, low-accuracy research\\ 
    \hline
    Barriers of \newline community \newline interaction & \textbullet\ Unestablished norms on how to use family trees \newline 
    \textbullet\ Conflicts between genealogists \newline
    \textbullet\ Recognizing and acknowledging expertise \newline
    \textbullet\ Privacy concerns\\ 
    \hline
    Challenges \newline in learning & \textbullet\ Locating learning resource that fits one's knowledge and needs \newline\textbullet\ Adopting appropriate research methods \newline
    \textbullet\ Leveraging genealogy \colored{software} systems\\
    \bottomrule
  \end{tabular}
  \caption{Summary of challenges faced by amateur genealogists to research effectively and conduct high quality research.
  }
  \label{tab:summary}
\end{table*}

\subsection{Technological Support for Research Quality}
Our findings on genealogy research are largely consistent with previous studies, including expert genealogists customizing research tools~\cite{duff2003list}, searching for local resources~\cite{yakel2007genealogists}, the requirement of history knowledge~\cite{friday2014learning}, and the importance of information organization and production~\cite{fulton2009quid}. We also confirmed that challenges resulting from inadequate genealogy research reported by \citet{willever2014family} persists a decade later, despite the emerging DNA testing and increasing availability of educational resources on the Internet. While our interviewees with professional backgrounds of lawyers and government analysts asserted that the “analytical skills” they acquired from previous careers are related to genealogical research, we also find common challenges with other fields of study, such as asymmetrical skepticism (confirmation bias)~\cite{marksteiner2011asymmetrical, shahid2022matches}, emotional proximity (connection with self-identity)~\cite{huang2015connected}, and digilantism (unsanctioned investigations of others)~\cite{starbird2014rumors, nhan2017digilantism}.

While genealogists praised the conveniences afforded by genealogy websites, we also found negative consequences of it. Specifically, we identified a high-productivity, low-accuracy information production phase at the very beginning of amateurs' research, in which complete beginners extensively utilize research hints provided by algorithms and build large family trees instantly. In this case, inexperienced genealogists and less reliable research hints amplify one other to exacerbate the situation. Unfortunately, such misinformation is extremely difficult to retract once is propagated. Even when amateurs are aware of their mistakes later, they may still keep these erroneous branches for future fixes, which may further hinder the effectiveness of the algorithm and mislead more users in a vicious cycle. We therefore encourage the development of recommendation algorithms that are sensitive to users' expertise and research quality which could suggest more reliable sources for amateurs.

Analyzing evidence is another task that amateurs struggle with. Genealogy websites may educate their users regarding the production of genealogical information, such as transcription, indexing, and algorithmic suggestions, to assist the evaluation of information. Besides bringing contextual information, we suggest that the platforms may also provide a canvas for cross-referencing multiple evidences and ancestors for correlating evidence and resolving conflict information, which is currently accomplished with customized spreadsheets by expert genealogists, but absent for amateurs. Automatic verification is also expected to help genealogists identify conflict information and avoid mistakes more effectively. Furthermore, since building hypothetical family trees is common for genealogists to explore other possibilities, more explicit differentiation of experimental trees with finished work might be introduced with a refined user interface.

Genealogists conduct original research as experts emphasize self-verifying on all clues, including system suggestions, family stories, and others' work. As a result, we observe that the evaluation of genealogical research is not solely based on the conclusion, but also the process used to reach it. Thorough documentation and organizing of the research is necessary for this goal, and experts also argue that it benefits genealogists as well by helping regulate their research. However, these tasks are time-consuming and often treated as secondary objectives to amateur genealogists, possibly due to the high cognitive demand of exhaustive information search and analysis. Therefore, genealogy platforms may consider automating these processes for users to reduce their overhead, but still prompt information such as current research goal to help regulate amateurs' research.

\subsection{Leveraging the Power of Community}
Our findings suggest that the online genealogist community occupies two types of platforms simultaneously. The first is genealogy websites, where genealogists build and organize family trees and directly collaborate with other users. The other is social media and online forums, in which genealogists find like-minded people, seek help, and learn from each other.

Genealogists may interact differently on these two types of platforms. On genealogy websites, amateurs are enthusiastic in sharing their family information with other genealogists, who are potential relatives and collaborators. On social media and open forums, users' privacy concerns may restrain sharing the same information with the public. Such differentiation also suggests that both types of platforms have their unique strengths. Genealogy websites can support collaborative problem solving, should such features be integrated into the system. For instance, a user could easily share a segment of their family tree and tag the ancestor with whom they face research challenges, and others who are familiar with the family, time period, and geographic location could step in with a more comprehensive view of the inquirer's research state, identify undiscovered mistakes, and provide advice accordingly. Online forums, on the other hand, are well-suited for socializing and asking more abstract questions with less risk of exposing personal information.

Implicit and explicit social interactions~\cite{serim2019explicating} may help genealogists establish community norms and avoid conflicts, but are often absent on genealogy websites. For example, featured family trees and ancestor profiles can be shown to amateur users to help them understand the expected research standard and how they should use family trees in their research. Expert genealogists may also benefit from this approach, as they may use featured family trees and profiles to showcase their research capability to amateurs, thus establishing community consensus on research expertise. To support more explicit social interaction, peer-review features such as rating and flagging of family trees would allow users to assess each others' research product and provide feedback more conveniently, which may also offset some of the challenges in direct communication. Finally, as previous studies~\cite{kim2021moderator, bagmar2022analyzing} suggested that chatbots could be effective for moderating online discussions and debates, similar technology could be deployed as a moderator to ease tensions and scaffold productive, evidence-based communication between genealogists who have different opinions, facilitating a more collaborative online environment.

\subsection{Computer Supported Andragogy}
Individuals' genealogy research often starts as a hobby, and we found amateurs' learning experience fits in well with the concept of andragogy, i.e., adult education and learning, as opposed to pedagogy, the education of dependent personalities such as children and students~\cite{knowles1973adult, forrest2006s}. Andragogy is characterized with problem-oriented, just-in-time learning, and leverages experience and knowledge of learners~\cite{forrest2006s}. Indeed, we found that many amateurs prefer learning during research. Their learning is self-directed, motivated by the challenge they are facing, and they need to identify the right learning resource that is suitable for their particular knowledge level.

We highlight a few characteristics of genealogy research and the genealogist community that make educating community members critical and challenging: Genealogy research is \textit{original} that every genealogist is going through the whole research process and expected to verify others' work before accepting it. It is \textit{personal} that family history is often associated with family stories and self-identity that not to be easily altered by others. The community is \textit{inclusive} that anybody, regardless of their genealogy research experience and literacy skills, is entitled to explore and publish their family history. Consequently, the overall quality of the collective product on genealogy websites is heavily dependent on the research capability of every individual. When the popularity of genealogy surged among millions of hobbyists, experts in this field faces significant challenges scaling up their traditional educational approaches.

Meanwhile, amateurs are not guaranteed to be exposed to the established research standards and resources with their self-directed learning.
Since amateurs rely on genealogy \colored{software} systems heavily at the beginning of their research, we believe that 
\colored{these} systems play a vital role in not only quality control, but also amateurs' education. On one hand, \colored{they} may enforce aspects of GPS to regulate amateurs' research practice. More importantly, it is critical to provide appropriate guidance and explanation throughout amateurs' research to help them aware more resources, understand the benefit of each research approach, and scaffold their learning of genealogy research. Machine learning based tutoring systems~\cite{spitzer2023ml} and large language models (LLM)~\cite{kasneci2023chatgpt, kung2023performance} could provide more scalable and user-friendly communication interface for amateur genealogists, while their effectiveness on genealogy research education needs further investigation.

As genealogists gain awareness of their knowledge gaps and seek to advance their research, the coming challenge is to identify the educational content that fits their need and knowledge level. Learnersourcing, an approach designed to support collective knowledge sharing among peer learners while maintaining their learning experiences~\cite{kim2015learnersourcing, glassman2016learnersourcing}, might be helpful for not only sharing insights on historical documents, but also recommending and synthesizing educational content among community members. Recognizing individuals' research experience in online forums could be challenging~\cite{willever2012tell}, while prior studies~\cite{murphy2009retina, piech2012modeling, zhang2023vizprog} suggesting that system logs may help characterize users' research behavior in classroom settings, which might also be applicable to genealogy \colored{software} systems. Future studies could be developed to evaluate users' genealogy expertise and challenges, and provide proper educational content accordingly.

\section{Limitations}\label{sec:limitations}

Identifying less experienced genealogists in an online community is not as straightforward as it sounds. We recruited participants who do not have credentials from the most accessible online community we can identify, expecting to reach varieties of genealogists in the wild. We acknowledge that novices who have not joined any community are not reachable with this recruiting approach, but we believe our participants' recall of their early research practices complemented some of the absent perspectives.

Recruiting from other communities and groups may introduce varying observations. However, our participants did not join Reddit or professional associations exclusively. They also shared activities in other online communities and groups, including Facebook groups, Discord, online genealogy study groups, local genealogy societies, and direct messaging through genealogy websites. Therefore, we believe our participants would be a valid representation of the broader online genealogist community.

Our study participants were all located in the United States at the time of their interviews and a majority of them are of white ethnicity. They might present a more U.S.-focused perspectives of family history research. Further studies could be conducted with genealogists from different geographical, cultural, and ethnic groups to expand our understanding of genealogy and family history studies worldwide.

\section{Conclusion}\label{sec:conclusion}
In this paper, we conducted an interview study to investigate genealogists' research process in detail, explored how amateurs' research might differ from established standards, and how it may introduce problems in an online, collaborative environment. We then explored social context of genealogist community, experts' and amateurs' perspectives on genealogy education, and their expectations for genealogy \colored{software} systems. We argued for the importance of technological support for quality control and educating newcomers of this community, and made design recommendations accordingly. As genealogists' independent research is aggregated by online genealogy platforms, this field offers a unique bridge between well-organized communities of practice and individuals' everyday information activity. More preciously, the genealogist community emphasizes explicit research standards and thorough documentation of research process. We believe that this community will provide HCI researchers a new lens through which to examine challenges such as supporting collaborative sensemaking process and information literacy education at scale in online communities.

\bibliographystyle{ACM-Reference-Format}
\bibliography{ref}

\appendix

\section{Interview Guide}\label{sec:interview}
Introduction:
\begin{enumerate}
\item How do you become interested in genealogy?
\begin{enumerate}
    \item What is your occupation? Is it related to genealogy?
\end{enumerate}
\item Can you tell me more about your research?
\begin{enumerate}
\item What is the primary focus (own family, others', research as a professional, etc.)?
\item How long have you been doing it?
\item How often are you involved in genealogy related activities? 
\item (Non-credential) How would you describe your expertise on genealogical research (novices, intermediate, expert, etc.)?
\end{enumerate}
\item What online genealogy communities do you participate in (Facebook, Reddit, Email lists, etc.)? How do you use them (seeking help, offering help,  seek information, etc.)?
\item (If tools include DNA) How do you use DNA test results for your genealogy research?\\
(If not) How do you feel about the DNA test? would you consider using them for your research?

\end{enumerate}
\noindent
Ongoing research:
\begin{enumerate}
\setcounter{enumi}{4}
\item How do you decide which question to start with or focus on in your research (what’s easiest, what’s most interesting, what the client asked for, etc.)?
\item Can you tell me about a time when you went down a research path or followed a hunch that you weren’t sure was correct or not? 
\item Have you ever explored several different answers to the same question simultaneously? 
\item How would you determine when a genealogy question is answered? 
\item How much time do you usually expect to solve a genealogy problem or decide to move to another topic?
\item How do you keep track of the progress before it’s solved?
\item Can you tell us about one time when you disagreed with other people’s conclusion? How did you resolve it?
\end{enumerate}
\noindent
Research barriers:
\begin{enumerate}
\setcounter{enumi}{11}
\item Can you tell me about a time when you ran into a brick wall?
\item What would you do to avoid or overcome brick walls?
\item (Non-credential)  How do you feel about credentials in genealogy research? Would you consider hiring a professional genealogist to research for you?
\item (Non-credential) Where do you seek help in your research? How do you tell someone is good at genealogy?
\end{enumerate}
\noindent
Information:
\begin{enumerate}
\setcounter{enumi}{15}
    \item How reliable do you think the records provided by websites like Ancestry.com and FamilySearch.org? How about the indexes?
    \item How reliable are the record hints provided in these websites? How do you use them in your research?
    \item How reliable do you think user-created family trees are? How do you use them in your research?
    \item (If negative) What do you think are the causes of this issue?
    \item (If negative) How do you identify and avoid unreliable information?
\end{enumerate}
\noindent
Sharing:
\begin{enumerate}
\setcounter{enumi}{20}
\item How do you share your research outcome with others?
\begin{enumerate}
    \item What steps would you take? In what formats? 
    \item What would you do to ensure the accuracy of your work?
\end{enumerate}
\end{enumerate}
\noindent
Education:
\begin{enumerate}
\setcounter{enumi}{21}
\item How did you learn how to do genealogy research? 
\item Are there any tools that you found could help newcomers?
\begin{enumerate}
    \item (Expert) How can you recognize a newcomer on the Internet?
\end{enumerate}
\item Can you imagine how we may help people become better genealogy researchers?
\end{enumerate}
\noindent
Changes:
\begin{enumerate}
\setcounter{enumi}{24}
    \item (For experienced genealogists) What do you think about genealogy have changed in the past decade?
\end{enumerate}
\noindent
Reflection:
\begin{enumerate}
\setcounter{enumi}{25}
    \item Any additional thoughts before we wrap up for today? Any other aspect of your research we should have asked but didn't?
    \item Anything you want to ask us?
\end{enumerate}

\end{document}